\documentclass[aps,twocolumn,nofootinbib,showpacs,floatfix]{revtex4}
\usepackage{graphics} 
\usepackage{amssymb,amsmath}
\usepackage{amsmath}
\usepackage{psfrag}
\usepackage{graphicx}
\newcommand{\ve}[1]{\mathbf{#1}}

\def\al{\alpha}

\def\ga{\gamma}

\def\si{\sigma}

\newcommand{\bm}[1]{\mbox{\boldmath $#1$}} 
 
\def\be{\begin{equation}} 
\def\ee{\end{equation}} 
\def\bea{\begin{eqnarray}} 
\def\eea{\end{eqnarray}}  
\def\bean{\begin{eqnarray*}} 
\def\eean{\end{eqnarray*}} 
\def\dd{\partial}

\def\bk{{\bf k}}

\def\br{{\bf r}}  
\def\ba{{\bf a}}

\def\bn{{\bf n}} 
\def\bu{{\bf u}}
\def\bR{{\bf R}}
\def\mD{{\mathcal D}}

\def\bb{{\bf b}}

\def\tmD{{\tilde\mD}}
\def\tbu{{\tilde\bu}}

\def\bse{\begin{subequations}}
\def\ese{\end{subequations}}

\def\da{{\dot a}}

\def\bF{{\mathbf F}}

\def\bm{\mathbf{m}}

\def\mP{\mathcal{P}}

\def\lsim{\raise 0.4ex\hbox{$<$}\kern -0.8em\lower 0.62ex\hbox{$\sim$}} 
\def\gsim{\raise 0.4ex\hbox{$>$}\kern -0.7em\lower 0.62ex\hbox{$\sim$}} 
\def\mP{\mathcal{P}}
\def\mQ{\mathcal{Q}}

\def\bk{\mathbf{k}}

\def\f0N{f_0^{(N)}}
\def\bec{\begin{center}}
\def\eec{\end{center}}

\def\bA{{\mathbf A}}

\def\mC{{\mathcal C}}
\def\bhk{{\bf \hat k}}

\begin{document} 
\title{Particle linear theory on a self-gravitating
perturbed cubic Bravais lattice}
\author{B. Marcos}   
\email{Bruno.Marcos@roma1.infn.it}
\affiliation{ ``E. Fermi'' Center, Via Panisperna 89 A, Compendio del 
Viminale, I-00184 Rome, Italy,\\
\& ISC-CNR, 
Via dei Taurini 19,
I-00185 Rome,
Italy.}

\begin{abstract}   
\begin{center}    
{\large\bf Abstract}
\end{center}    
Discreteness effects are a source of uncontrolled systematic errors of
N-body simulations, which are used to compute the evolution of a
self-gravitating fluid. We have already developed the so-called {\em
``Particle Linear Theory''} (PLT), which describes the evolution of
the position of self-gravitating particles located on a perturbed
simple cubic lattice. It is the discrete analogue of the well-known
(Lagrangian) linear theory of a self-gravitating fluid. Comparing both
theories permits to quantify precisely discreteness effects in the
linear regime. It is useful to develop the PLT also for other
perturbed lattices because they represent different discretizations of
the same continuous system. In this paper we detail how to implement
the PLT for perturbed cubic Bravais lattices (simple, body and
face-centered) in a cubic simulation box. As an application, we will
study the discreteness effects --- in the linear regime --- of N-body
simulations for which initial conditions have been set-up using these
different lattices.
\end{abstract}    
\maketitle   
\date{today}  

\twocolumngrid

\section{Introduction}

An important problem in cosmology is the formation of the large scale
structure. The key process involved is the gravitational clustering of
collisionless dark matter, which is considered to be well described as
a self-gravitating fluid for a wide range of scales
(e.g. \cite{peebles_80}). The complexity of these fluid equations
(coupled with gravity) makes impossible to compute an analytical
solution. There are therefore two common approaches to attack the
problem: {\em (i)} a perturbative expansion in the density contrast
$\delta(\br)=\rho(\br)/\rho_0-1$ (where $\rho(\br)$ is the local
density and $\rho_0$ its space average), valid only at early times (or
for scales in which the density contrast averaged over such scales is
smaller than one) and {\em (ii)} N-body simulation, in which the fluid
is discretized into particles (N-bodies) and then the evolution of the
system computed applying simple gravity.

N-body simulations are used to compute the evolution in the highly
non-linear regime. A basic problem of this method is that there is no
theory on the discreteness effects due to the use of a finite number
$N$ of particles
(e.g. \cite{melott_90,kuhlman_96,melott_97,splinter_98,heitmann_04,diemand_04}). Generally,
tests varying $N$ shows a ``convergence'' of the simulations.
However, it is difficult to infer how well this convergence has been
achieved because of the lack of framework to refer to. For example, it
is not known the dependence of the discreteness error with $N$. If the
convergence is slow, numerical tests can indeed appear to converge
when actually convergence has not been achieved (see
e.g. \cite{power_03,diemand_04}).

In \cite{joyce_04} and subsequently \cite{joyce_05,marcos_06,joyce_07}
we have started to develop a program to precisely fill this gap. We
have developed a framework which allows us to calculate the evolution
--- in the linear regime --- of a system of self-interacting
particles. This is the discrete counterpart of the well-known {\em
Fluid Linear Theory} (hereafter FLT), and we called it {\em Particle
Linear Theory} (hereafter PLT).  We have shown that the fluid limit of
the PLT is well defined and indeed it is the FLT. We have shown also
how to quantify, in an essentially analytic way, the discreteness
effects, with arbitrarily large precision. Moreover, availability of
analytical results permits to evaluate the discreteness effects in the
limit of infinite realizations. It avoids, in the computation of
statistical quantities, the use of any statistical estimator and thus
its subsequent and problematic noise. One of our conclusions was that,
for the set-up of the initial conditions (IC), the body centered cubic
lattice could be a better choice than the simple cubic (sc) one,
because it might produce less discreteness effects. We will see in
this paper that it is indeed the case in this context of linear
theory.

Moreover, another important motivation of this paper is the study of
the discreteness effects in the $non-perturbative$ regime. In the
forthcoming paper \cite{marcos_07} we sample the {\em same} continuous
field using different lattices and then evolve them using N-body
simulations. The differences between the result of these simulations
give an estimate of the lower bound of the discreteness effects in the
{\em non-perturbative regime}. Because these differences are small ---
typically of order of a few percent in the power spectrum for times
and scales relevant to cosmological simulations ---, an implementation
of the PLT for these lattices is an essential tool to check that these
differences are actually discreteness effects and not numerical
errors, finite-size effects, estimator-related errors, etc.

In this paper we present the PLT method applied to any cubic Bravais
lattice, i.e., to a simple cubic (sc), body-centered (bcc) and
face-centered (fcc) lattices. In the first section we give an summary
of the PLT. Further details can be found in \cite{marcos_06}. In the
following section, we explicitly give the details of the PLT for a sc,
bcc and fcc lattices. We use Fast Fourier Transform techniques, in a
{\em cubic box}, which is a-priori non trivial. In the last section,
we present some applications of the method, comparing discreteness
effects using a perturbed sc, bcc or fcc lattice to set-up the IC. It
is a generalization to a bcc and fcc lattices of the work presented in
\cite{joyce_07}.

\section{Linearization of gravity on a perturbed lattice}
\label{Grav-lat}

In this section we present a summary of the general method we have
developed in \cite{joyce_05,marcos_06} to calculate the evolution of
self-gravitating particles perturbed off a perfect lattice. 

Let us consider a parallelepiped of volume $V$ with $N$ lattice sites,
which are generated combining linearly the three primitive lattice
vectors $\ba_1$, $\ba_2$ and $\ba_3$:
\be
\label{bravais}
\bR=\bR(n_1,n_2,n_3)=\ell(n_1\ba_1+n_2\ba_2+n_3\ba_3),
\ee
where 
\be
\label{number-real}
n_i\in[0,N_i-1]\cap \mathbb Z
\ee
and $\ell$ is the typical ``lattice spacing''\footnote{For a sc
lattice $\ell$ is the actual lattice spacing while this is not true in
the bcc and fcc case, because all the lattice sites are not at the
same distance each one from another.} (we have chosen $\ba_i$ to be
dimensionless). The total number of particles in the system is
$N=N_1N_2N_3$ and the box a parallelepiped with sides
$\bA=\{\bA_1,\bA_2,\bA_3\}$, where $\bA_i=N_i\ba_i$.

We perform a displacement of the particles about their lattice
position $\bR$ and we write their new position $\br(t)$ as:
\be
\label{lagrangian}
\br(t)=\bR+\bu(\bR,t),
\ee
which will evolve under the effect of gravity and where $\bu(\bR,t)$
is a {\em displacement field} evaluated at the lattice positions.

\subsection{Definition and linearization of the gravitational force}

In order to have a translationally invariant system\footnote{This is
not to have any privileged point in the system.}  we take periodic
boundary conditions.  We use the {\em method of replicas} to compute
the gravitational force. It consists in calculating the force not only
considering the particles in the box of volume $V$ but also all its
{\em images}, i.e., an infinite number of copies of the system. This
is a standard scheme in cosmological N-body simulations to evaluate
the force (see e.g. \cite{hockney_88}). For a well defined
gravitational force in the infinite volume limit, it is necessary to
introduce a neutralizing background which, in cosmology, is naturally
introduced in the context of an expanding universe (see
e.g. \cite{peebles_80}).

The gravitational force is linearized by expanding in Taylor series at
linear order in the variable $\bu(\bR,t)$ about the lattice position
$\bR$ (for more details see e.g. \cite{marcos_06}).  It is convenient
to use of the {\em dynamical matrix} $\mD(\bR)$ to express the
linearized force:
\be
\label{dynamical-eq}
\bF(\br)=\sum_{\bR'}\mD(\bR-\bR')\bu(\bR').
\ee
The expression of the dynamical matrix  for a generic
interaction potential $v(r)$ is \cite{marcos_06}:
 \bse
\label{expr_D}
\begin{align}
\label{expr_D_nonzero}
\mD_{\mu\nu}(\bR\ne\mathbf 0)&=\dd_\mu \dd_\nu w(\bR)\\
\label{expr_D_zero}
\mD_{\mu\nu}(\bR=\mathbf 0)&=-\sum_{\bR'\neq 0} \dd_\mu \dd_\nu w(\bR')
\end{align}
\ese
where
\be
\dd_\mu \dd_\nu w(\br_0)=\left[\frac{\partial^2\,w (\br)}{\partial r_\mu \partial r_\nu}\right]_{\br=\br_0}
\ee
and $w(\br)$ is the periodic function defined as
\be
\label{w-def}
w(\br)=\sum_{\bn}v(\br+\bn\cdot \bA) , \ee i.e., the potential due to
a single particle and all its copies.  For the gravity force, we have
$v(\ve r)=-Gm/r$ and Eq.~\eqref{w-def} is implicitly understood to be
regularized by the addition of a uniform negative background. However,
the sum \eqref{w-def} is numerically slowly convergent (it is
necessary to sum over a huge number of replicas). To speed-up the
computation we use the standard Ewald method, which consists in
dividing the sum in a short range part and a long range one
introducing a damping function $\mC$:
\begin{equation}
\begin{split}
\label{pot_damping}
w(\br)&= \sum_{\bn} v(\br+\bn\cdot\bA){\mathcal C}(|\br+\bn\cdot \bA|,\alpha)\\
&+\sum_{\bn} v(\br+\bn\cdot\bA)[1-{\mathcal C}(|\br+\bn\cdot\bA|,\alpha)], 
\end{split}
\end{equation}
where $\al$ is a damping parameter from which the result is independent. A common choice for a $1/r$ potential is
\be 
{\mathcal C}(|\br|,\alpha)={\mathrm{erfc}}(\alpha|\br|).  
\ee 
The expression for the function $w$ is then: 
\be
\label{pot_ewald_div}
w(\br)=w^{(r)}(\br)+w^{(k)}(\br)
\ee
and
\bse
\label{pot_ewald}
\begin{align}
\label{pot_ewald_r}
w^{(r)}(\br)&=-Gm\sum_{\bn}\frac{1}{|\br+\bn\cdot \bA|}{\mathrm{erfc}}(\alpha|\br+\bn\cdot \bA|),\\
\label{pot_ewald_k}
w^{(k)}(\br)&=-Gm\frac{4\pi}{V_B}\sum_{\bk\neq \ve
0}\frac{1}{|\bk|^2}\exp\left(-\frac{|\bk|^2}{4\alpha^2}\right)
\cos\left[\bk\cdot\br\right].
\end{align}
\ese
The Fourier vectors $\bk$ are generated combining linearly the
(dimensionless) primitive vectors in reciprocal space $\bb_i$
\be
\label{k-vectors}
\bk\ell=m_1\frac{\bb_1}{N_1}+m_2\frac{\bb_2}{N_2}+m_3\frac{\bb_3}{N_3}
\ee
where $m_i$ are integers and
\be
\label{ab}
\ba_i\cdot\bb_j=2\pi\delta_{ij}
\ee
($\delta_{ij}$ is the Kronecker delta). We define the {\em Nyquist frequency} as
\be
\label{nyquist}
k_N=\frac{\pi}{\ell}.
\ee
It is simple to show (e.g \cite{marcos_06}) that the term
$\bk=\mathbf0$ is not included in the sum \eqref{pot_ewald_k} due to
the presence of the neutralizing background (or the space expansion in
the cosmological context). An explicit expression of the dynamical
matrix is given in App.~\ref{ewald-app}.

\subsection{Dynamical equations}
\label{dynamical}
For simplicity we will consider a matter-dominated universe with zero
cosmological constant (Einstein-deSitter, hereafter EdS)\footnote{For
a static non-expanding universe see \cite{marcos_06}.}. This is a very
good approximation for the currently most favored $\Lambda$CDM
cosmological model for the times in which PLT is a good approximation
(i.e. before shell-crossing), considering the typical red-shifts in
which the simulations are started. The evolution of the displacement
field $\bu(\bR,t)$ is given by the equation
\be
\label{eq-motion-exp}
\ddot\bu(\bR,t)=-2\frac{\da}{a}\dot\bu(\bR,t)+\frac{1}{a^3}\sum_{\bR'}^N\mD(\bR-\bR')\bu(\bR',t),
\ee
where $a(t)$ is the scale factor and the (double) dots mean (double)
derivative with respect to time. From Bloch theorem it is possible to
diagonalize Eq.~\eqref{eq-motion-exp} in real space using the following
combination of plane waves:
\be
\label{eigenvector}
\bu(\bR,t)=\frac{1}{N}\sum_{\bk}\tbu(\bk,t)e^{i\bk\cdot\bR}, 
\ee
where the sum is restricted to the {\em first Brillouin zone}
(hereafter FBZ), i.e., by the set of the $N$ vectors $\bk$\footnote{It is
simple to show (e.g. \cite{pines_63}) that a periodic lattice with $N$
particles has $N$ associated independent vectors $\bk$.} with smaller
modulus. These symmetrically lie  around $\bk=0$\footnote{The FBZ is not
in general symmetric about $\bk=\mathbf 0$ but this is the case for a
{\em cubic} Bravais lattice because of the symmetries of the
lattice.}. We denote $\tbu(\bk,t)$ as the Fourier transform (hereafter
FT) on the lattice of $\bu(\bR,t)$
\be
\label{eigenvector-inv}
\tbu(\bk,t)=\sum_{\bR}\bu(\bR,t)e^{-i\bk\cdot\bR}, 
\ee 
where the sum is restricted to the simulation box (i.e. without
considering the replicas). Using Eqs.~\eqref{eq-motion-exp} and
\eqref{eigenvector} we obtain the $3\times3$ eigenvalue problem
\be
\label{eq-motion-k}
\ddot\tbu(\bk,t)=\tilde{\mathcal D}(\bk) \bu(\bk,t),
\ee
where $\tilde{\mathcal D}(\bk)$ is defined analogously to $\tbu(\bk,t)$. We can easily diagonalize (numerically) Eq.~\eqref{eq-motion-k}, obtaining for each $\bk$ the eigenvalue equation
\be
\label{eigen_equation}
\tmD(\bk)\mathbf{\hat e}_n(\bk)=4\pi G\rho_0\, \varepsilon(\bk)\mathbf{\hat e}_n(\bk),
\ee
where $\rho_0$ is the average mass density $\rho_0=n/V$ and $\varepsilon(\bk)$ the normalized eigenvalues of the dynamical matrix $\tmD(\bk)$. We can decompose each mode $\tbu(\bk,t)$ in the basis $\{\mathbf{\hat e}_n(\bk),\,n=1,2,3\}$ as
\be
\label{transf-eigen}
\tbu(\bk,t)=\sum_{n=1}^3\mathbf{\hat e}_n(\bk) f_n(\bk,t).
\ee
Using Eqs.~\eqref{eq-motion-k}, \eqref{eigen_equation} and
\eqref{transf-eigen} we get the following equation for the
coefficients $f_n(\bk,t)$:
\be
\label{modes-exp}
\ddot f_n(\bk,t)+2\frac{\da}{a}\dot f_n(\bk,t)=\frac{4\pi G\rho_0\, \varepsilon(\bk)}{a^3}f_n(\bk,t).
\ee
Depending on the sign of $\varepsilon(\bk)$, we obtain two classes of
solutions $U_n(\bk,t)$ and $V_n(\bk,t)$, which are given in App. \ref{mode-sol}.

\subsection{Evolution of the power spectrum}

Usually, we are not interested in the position of each particle but in
some global statistical quantities. In this paper, we will focus on the
power spectrum (hereafter PS), defined as
\be
\label{PS_def}
P(\bk)=\lim_{V\to\infty}\frac{\langle \delta\tilde\rho(\bk)\delta\tilde\rho^*(\bk)\rangle}{V},
\ee
where $\delta\tilde\rho(\bk)$ is the FT of the density contrast
$\delta\rho(\br)=\rho(\br)-\rho_0$ (we assume statistical
homogeneity). It is possible to show that for a small value of the
displacement $|\bu(\bR,t)|\ll\ell$, the PS of a perturbed
lattice can be written as \cite{gabrielli_04,joyce_04}
\be
\label{PS-smalld}
P(\bk,t)\approx k_\mu k_\nu \tilde g_{\mu\nu}(\bk,t),
\ee
where 
\be
\tilde g_{\mu\nu}(\bk)=\lim_{V\to\infty}\frac{\langle \tilde u_\mu(\bk)\tilde u_\nu^*(\bk)\rangle}{V}.
\ee
Setting-up the IC at $t=t_0$ in the canonical way using the Zeldovich
approximation is equivalent to set (e.g. \cite{joyce_04})
\be
\label{g-ZA}
\tilde g_{\mu\nu}(\bk,t_0)=\hat \bk_\mu \hat\bk_\nu \tilde g(\bk,t_0).
\ee
 Using Eqs.~\eqref{eigen_evol}, \eqref{evol_operators}, \eqref
{PS-smalld} and \eqref{g-ZA} we get:
\be
\label{P-evol}
P(\bk,t)\approx A_P^2(\bk,t)P(\bk,t_0),
\ee
where
\be
\label{AofP}
A_P(\bk,t)=\sum_{\mu,\nu}\hat\bk_\mu\hat\bk_\nu {\cal A}_{\mu\nu}(\bk,t)
\ee
and (for an EdS universe) \cite{marcos_06}
\be
\label{amunu}
{\cal A}_{\mu\nu}(\bk,t)=\sum_{n=1}^3 \left[U_n(\bk,t)+\frac{2}{3t_0}V_n(\bk,t)\right](\hat {\mathbf e}_n)_\mu (\hat {\mathbf e}_n)_\nu.
\ee

\section{Diagonalization of the dynamical matrix}

In this section we describe step-by-step how to diagonalize the
dynamical matrix. 

\subsection{Generation of the real space lattice}

In general, N-body simulations are performed in a {\em cubic} box,
using a perturbed lattice as initial conditions. Therefore, to fill
the simulation box in an uniform way, the number of particles cannot
be arbitrary. In the case of a sc lattice, the number of points should
be $N=N_{{\mathrm sc}}^3$ (with $N_{\mathrm sc}$ an integer), for a
bcc one $N=(N_{\rm bcc}/2)^3$ and for an fcc one $N=(N_{\rm fcc}/4)^3$
(where $N_{{\rm bcc}}$ and $N_{{\rm fcc}}$ are also integers).

Note
that the real space vectors $\bR$, generated using
Eq.~\eqref{bravais}, lie, in general, in a {\em parallelepiped} box, with sides $\{\bA_1,\bA_2,\bA_3\}$. Note that it is
necessary to generate the real space vectors in this way [i.e. using
Eq.~\eqref{bravais} and \eqref{number-real}] in order to use the
technique of Fast Fourier Transform (FFT) as we will see in section
\ref{FFT}. We have therefore to translate the $\bR$ vectors into a
cube using a operation which leaves unchanged the dynamics of the
system. It is simple to show that the equation of motion
~\eqref{eq-motion-exp} is invariant under the transformation
\be
\label{translate-cubic}
\bR\longrightarrow \bR+\sum_{i=1}^{3} n'_i\bA_i,
\ee
(where $n'_i$ are integers). We can, then, choose three primitive
lattice vectors $\{\ba_i ,\,i=1,2,3\}$ and the number of particles
$N_i$ associated with each primitive lattice vector (compatible with
the total number of particles) which, using
Eq.~\eqref{translate-cubic}, translate all the lattice sites into a
cube. This is not trivial and does not work for any combination of
primitive lattice vectors and number of particles in each direction
(compatible with the total number of particles). We give in
Table~\ref{lattice-table} a set of primitive lattice vectors and in
Table~\ref{number-table} the particle number associated with them (for a
total of $N$ particles) for a sc, bcc and fcc lattices which fulfill
the above requirements.

\begin{table}
\begin{tabular}{|c||c|c|c|}
\hline
& $\ba_1$ & $\ba_2$ & $\ba_3$  \\
\hline
\hline
sc & $[1,0,0]$ & $[0,1,0]]$ & $[0,0,1]$ \\
\hline
bcc & $\ [1,0,0]$ & $\ [0,1,0]]$ & $\frac{1}{2}\ [1,1,1]$ \\
\hline
fcc & $\frac{1}{2} \ [0,1,1]$ & $\frac{1}{2}\ [1,0,1]]$ & $\ [0,1,0]$ \\
\hline
\end{tabular}
\caption{Lattice vectors  for the different kind of cubic Bravais lattices.
 \label{lattice-table}}
\end{table}

\begin{table}
\begin{tabular}{|c||c|c|c|}
\hline
 & $N_1$ & $N_2$ & $N_3$ \\
\hline
\hline
sc  & $N^{1/3}$ & $N^{1/3}$ & $N^{1/3}$\\
\hline
bcc  & $\left(\frac{N}{2}\right)^{1/3}$ & $\left(\frac{N}{2}\right)^{1/3}$ & $2\left(\frac{N}{2}\right)^{1/3}$\\
\hline
fcc  & $\left(\frac{N}{4}\right)^{1/3}$ & $\left(\frac{N}{4}\right)^{1/3}$ & $4\left(\frac{N}{4}\right)^{1/3}$ \\
\hline
\end{tabular}
\caption{Associated number particles with the lattices vectors listed in Table~\ref{lattice-table} for the different kind of cubic Bravais lattices.
 \label{number-table}}
\end{table}

\subsection{Generation of vectors in reciprocal space in the FBZ}

Given the primitive lattice vectors $\ba_i$, the primitive vectors in
reciprocal space are univocally defined by Eq.~\eqref{ab}. The
basis we have used to generate the lattices is given in Table
\ref{lattice-table} and the corresponding primitive reciprocal vectors
are listed in Table \ref{kvectors-table}.
\begin{table}
\begin{tabular}{|c||c|c|c|}
\hline
& $\bb_1$ & $\bb_2$ & $\bb_3$ \\
\hline
\hline
sc & $2\pi[1,0,0]$ & $2\pi[0,1,0]]$ & $2\pi[0,0,1]$ \\
\hline
bcc & $2\pi[1,0,-1]$ & $2\pi[0,1,-1]]$ & $4\pi[0,0,1]$\\
\hline
fcc & $4\pi[-1,0,1]$ & $4\pi[1,0,0]]$ & $2\pi[1,1,-1]$\\
\hline
\end{tabular}
\caption{Reciprocal vectors for the different lattices.
 \label{kvectors-table}}
\end{table} 
 The reciprocal vectors are generated using Eq.~\eqref{k-vectors}
where $m_i$ are the same integers as the ones used to generate the
$\bR$ vectors, i.e.,
\be
\label{generate-k}
m_i\in[0,N_i-1]\cap \mathbb Z.
\ee
It is necessary, in order to use FFT techniques, to
generate the reciprocal vectors in this way, as we will see in
section \ref{FFT}.

However, all the $\bk$ vectors used in the computation of the
evolution of the particle position must lie in the FBZ (see section
\ref{dynamical}) but, in general, those generated using
Eqs.~\eqref{k-vectors} and \eqref{generate-k} do not. We can translate
the reciprocal vectors into the FBZ using the transformation which
leads Eq.~\eqref{eq-motion-k} invariant
\be
\label{get-Brillouin}
\bk\longrightarrow \bk+\sum_{i=1}^{3} m'_i\bb_i,
\ee
where $m'_i$ are some appropriate integers.

One can obtain a complete set of $N$ $\bk$ vectors which are in the
FBZ, in the following way: compute a set of candidate vectors to lie
in the FBZ with Eqs.~\eqref{k-vectors} and \eqref{get-Brillouin}. To
select those which are in the FBZ, it is not efficient to consider the
$N$ vectors with smaller modulus because it is an ${\cal O}(N^2)$
operation. The computation time for this can be prohibitive for large
$N$. It is much better to construct geometrically the shape of the FBZ
by considering some point of the reciprocal space (namely $\bb=\mathbf
0$) and then drawing the perpendicular bisector planes of the
translation vectors from the chosen center to the nearest sites of the
reciprocal lattice. In Table \ref{FBZ-table}, we give the normal
vector of this plane, with modulus equal to their closest distance to
the center $\bk=\mathbf 0$. The FBZ of the sc lattice is a cube of
side $2\pi/\ell$, the one of the bcc lattice a rhombic dodecahedron
and the one of the fcc lattice a cuboctahedron. Then, we select the
$\bk$ vectors which are enclosed between these planes. This is an
essentially ${\cal O}(N)$ operation.

\begin{table}
\begin{tabular}{|c|c|c|}
\hline
sc & bcc & fcc \\
\hline
$\frac{\pi}{\ell}[\pm 1,0,0]$ &  $\frac{\pi}{\ell}[0,\pm1,\pm1]$ &  $\frac{2\pi}{\ell}[\pm1,0,0]$ \\
$\frac{\pi}{\ell}[0,\pm 1,0]$ & $\frac{\pi}{\ell}[\pm1,0,\pm1]$ & $\frac{2\pi}{\ell}[0,\pm1,0]$ \\
$\frac{\pi}{\ell}[0,0,\pm 1]$ & $\frac{\pi}{\ell}[\pm1,\pm1,0]$  & $\frac{2\pi}{\ell}[0,0,\pm1]$ \\
& & $\frac{\pi}{\ell}[\pm1,\pm1,\pm1]$ \\
$2+2+2$ vectors & $4+4+4$ vectors & $2+2+2+8$ vectors\\
\hline
\end{tabular}
\caption{Normal vectors which define the FBZ of the bcc and fcc lattices.
 \label{FBZ-table}}
\end{table}

\subsection{Fast Fourier Transform}
\label{FFT}
In this section, we will carry out the FFT of some
quantity defined on the lattice as, e.g., the dynamical matrix
\be
\label{FT-dynamical}
\tmD(\bk)=\sum_{\bR} \mD(\bR) e^{i\bk\cdot\bR},
\ee
where $\bR$ is restricted to the simulation box. Equation
\eqref{FT-dynamical} involves an ${\cal O}(N^2)$ operations (an
$N$-term sum for each of the $N$ $\bk$ vectors). However, using the
so-called {\em Fast Fourier Transform} (FFT) technique, it is possible
to reduce the number of operations --- exploiting the symmetries of
the FT --- to only $N\ln_2 N$ operations. We give a brief summary of
how the FFT works in App.~\ref{app-FFT}. By using it, we can speed-up
greatly the computation of the FTs of the dynamical matrix and the
displacement field. Using Eqs.~\eqref{bravais}, \eqref{k-vectors} and
\eqref{ab} we can write Eq.~\eqref{FT-dynamical} as
\be
\label{FFT-indices}
\tmD_\bm=\sum_{\bn} \mD_n \exp\left[2\pi i \left(\frac{n_1m_1}{N_1}+\frac{n_2m_2}{N_2}+\frac{n_3m_3}{N_3}\right)\right],
\ee
where the indices $\bn$ and $\bm$ labels the $\bR$ and $\bk$ vectors
respectively. These are the same triplets of integers which have been
used in Eqs.~\eqref{bravais} and \eqref{k-vectors} respectively. Note
that Eq.~\eqref{FFT-indices} is a three-dimensional FT, i.e., three
embedded one-dimensional FT as the one of Eq.~\eqref{FT-1d}, with the
same running of indices [see Eqs.~\eqref{number-real} and
\eqref{generate-k}]. It is then straightforward to compute the FT
\eqref{FFT-indices} using any standard FFT routine. Note that each
$\bR$ vector should be associated in Eq.~\eqref{FFT-indices} with the
indices $[n_1,n_2,n_3]$ with which it has been generated using
Eq.~\eqref{bravais}, and not those that would correspond to their actual
position in the cubic box after being applied the transformation
\eqref{translate-cubic}. The same observation holds for the $\bk$
vectors, whose indices correspond to those used generating them with
Eq.~\eqref{k-vectors}.

There exists a great number of publicly available very competitive FFT
routines. We have used the {\em Fastest Fourier Transform in the West}
(FFTW) \cite{fftw}, since it can be used for any number of particles (and not
only powers of two).

\subsection{Spectrum of eigenvalues of an sc, bcc and fcc lattices}
\begin{figure}
\includegraphics[width=0.48\textwidth]{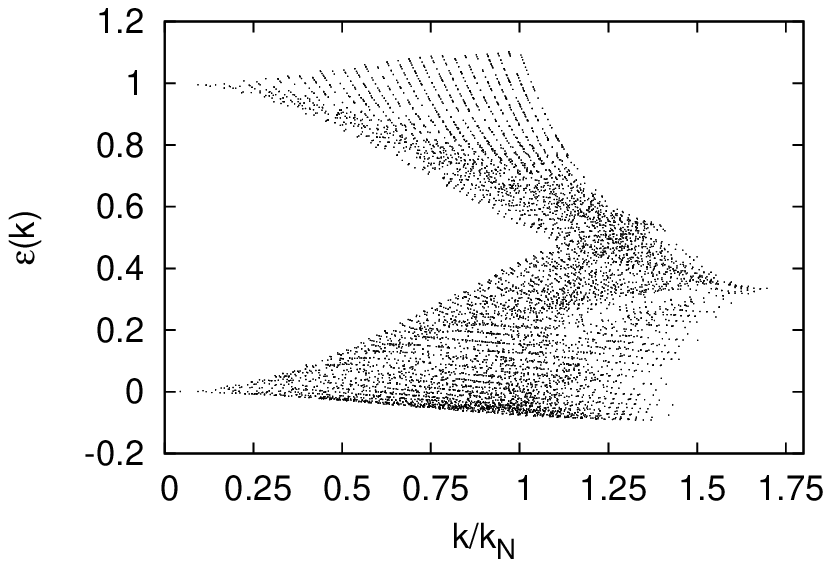}
\\
\includegraphics[width=0.48\textwidth]{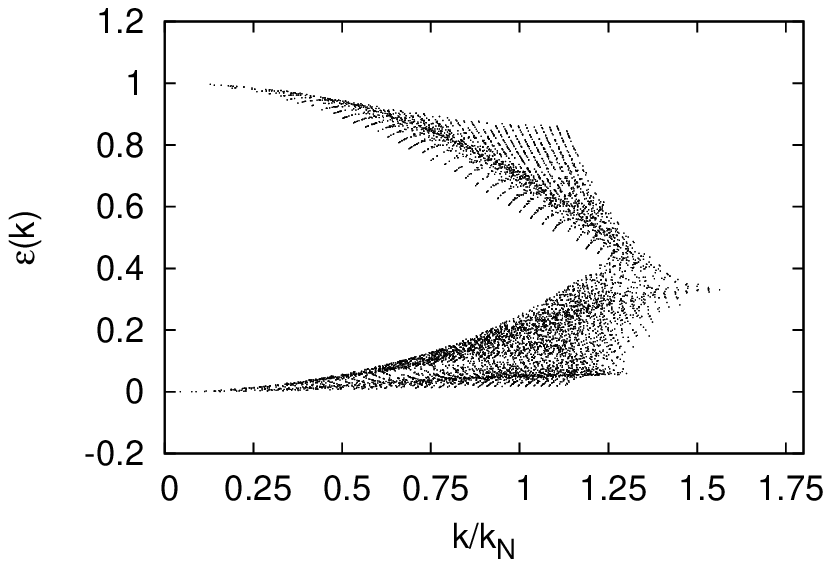}
\\
\includegraphics[width=0.48\textwidth]{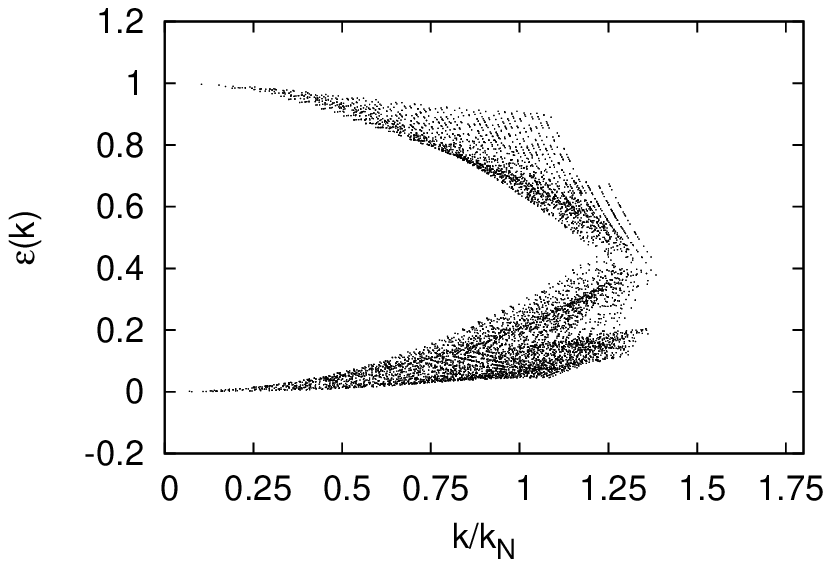}
\caption{Normalized spectrum of eigenvalues of (from top to bottom) a
sc, bcc and fcc in function of the wavector normalized to the Nyquist
frequency of the sc lattice defined in Eq.~\eqref{nyquist}. We have performed a sampling taking $1\%$ of the
points.
 \label{spectrum}}
\end{figure}
As an application of these techniques, we show in Fig.~\ref{spectrum}
the spectrum of eigenvalues corresponding to a sc, bcc and fcc
lattice. We use a cubic box and $N_{sc}=64$ for the sc lattice. For a comparison with the other lattices, we need
$N_{\rm bcc}=51$ and $N_{\rm fcc}=40$.
The three lattice presents the same branch structure (for further
discussion see \cite{marcos_06}): {\em (i)} an {\em optical branch},
with eigenvalues $\varepsilon(\bk\ell\to\mathbf 0)=1$ and eigenvector
polarized parallel to $\bk$ (in the same limit) and {\em (ii)} two
{\em acoustic branches} with normalized eigenvalues
$\varepsilon(\bk\ell\to\mathbf 0)=0$ and polarized in the plane
transverse to $\bk$ (in the same limit). We also see that, as
anticipated in \cite{marcos_06}, the spectrum of the bcc and fcc
lattices does not present negative nor eigenvalues with
$\varepsilon(\bk)>1$.

\section{Discreteness effects in a bcc, fcc and sc lattice}
\label{discreteness}

In this section, we will apply the method described above to compare
the discreteness effects when using different lattices to set-up the
IC, i.e., different discretizations of the same initial density
field. To do that, we compare FLT with PLT for the three lattices
considered. We will study two different effects: the change in the
amplification of the PS and the breaking of isotropy. A more detailed
study of discreteness effects in the linear regime for a sc lattice
can be found in \cite{joyce_07}.

In the fluid limit the evolution of the PS is given by the well-known
FLT (e.g.\cite{peebles_80,marcos_06}):
\be
P^{\rm fluid}(k,t)=a^2(t) P(k,t_0)
\ee
[we consider an initial PS which is statistically isotropic and we have used
that $a(t_0)=1$, see Eq.~\eqref{a-EdS}]. The evolution using PLT,
is given by
Eqs.~\eqref{P-evol} and \eqref{AofP}. We set up the IC by using the ZA approximation. 

\begin{figure}
\includegraphics[width=0.48\textwidth]{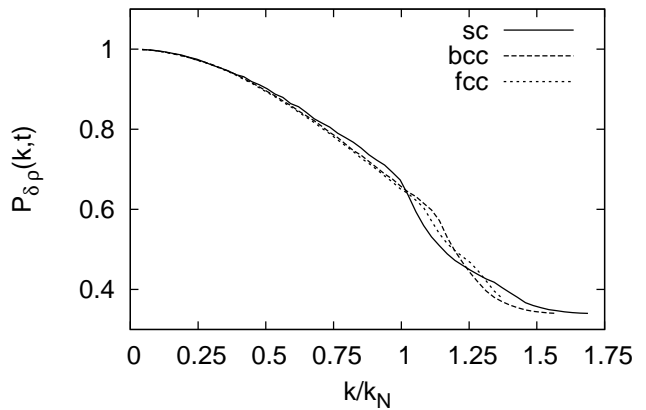}
\caption{Amplification of the PS averaged over bins normalized to the
fluid (FLT) amplification. The wavevectors have been normalized to the
Nyquist frequency of the sc lattice (see the text for more details).
 \label{amplication-av}}
\end{figure}

To characterize the effect of
the discreteness in the amplification of the PS we define the quantity
\be
\label{ampli-PS}
P_{\delta\rho}(\bk,t)=\
\frac{P(\bk,t)}{P^{\rm fluid}(k,t)},
\ee
which is the amplification of the PS calculated with PLT normalized by
FLT.  In Fig.~\ref{amplication-av} we show the amplification of the PS
at $a(t)=5$ predicted by PLT, normalized to the fluid
amplification. We have averaged over $60$ bins centered in $|\bk|$
with amplitude $|\bk|\pm|\Delta\bk|$ with $|\Delta\bk|\approx 0.92$.
We see that the evolution of the sc lattice is slightly closer to the
fluid one ($A_P^2(\bk,t)/a(t)^2=1$) for $k\lesssim k_N$ than the bcc
and fcc lattices. This is not surprising, looking at the form of the
spectrum of eigenvalues of the different lattices shown in
Fig.~\ref{spectrum}. The amplification of {\em each single} mode for
$k\ll k_N$ of the PS is related essentially with the shape of the {\em
optical branch} of the spectrum of eigenvalues. There are some
eigenvalues in the sc lattice with $\varepsilon(\bk)>1$ which
compensate, averaging over bins of same $|\bk|$, the largest part of
eigenvalues with $\varepsilon(\bk)<1$. However, for $k\gtrsim k_N$,
the evolution of the sc lattice is farther than the other two from the
fluid evolution. In fact, the modes with $\varepsilon(\bk)>1$ do not
exist anymore for $k>k_N$. Therefore, looking at the amplification of
the PS, we can say that the sc lattice is slightly closer to the fluid
evolution for $k\lesssim k_N$. However, as we will see below, the {\em
anisotropy} introduced by the sc lattice is much larger than the one
introduced by the bcc or fcc lattice.

\begin{figure}
\includegraphics[width=0.48\textwidth]{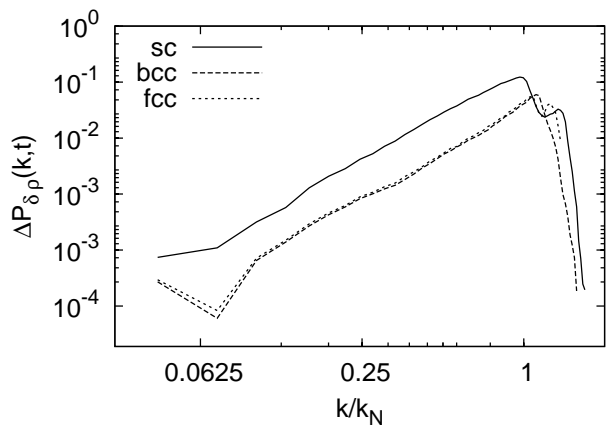}
\caption{Normalized dispersion of the PS amplification $\Delta
P_{\delta \rho}(k,t)$ averaged over bins of the same modulus of $\bk$.
 \label{dispersion}}
\end{figure}

 Let us consider the {\em
normalized dispersion} of the amplification of the PS, defined as in
\cite{joyce_07}
\be
\label{ddrho}
\Delta P_{\delta \rho}(k,t) =
\left(\frac{\overline{P_{\delta \rho}^2} (k,t) - 
\overline{P_{\delta \rho}}^2 (k,t)}
{\overline{P_{\delta \rho}}^2 (k,t)}
\right)^{1/2}
\ee
where the average, of any function ${X}(\bk, t)$ on the reciprocal
lattice, is defined as 
\be \overline{X} (k,t) = \frac{1}{N_k}
\sum_{\bk , |\bk|=k} {X}(\bk, t), \ee 
$N_k$ being the number of eigenmodes at a given $k$. This quantity
gives a measure of the anisotropy of the PS amplification. In a
system which respects the isotropy of a fluid, the amplification of
plane waves with the same corresponding wavevector, but different
direction, should be the same. In Fig.~\ref{dispersion} we show that
the sc presents a dispersion which is about an order of magnitude
larger than the one of the bcc or fcc, which is very similar. The
behavior of the dispersion $\Delta P_{\delta \rho}(k,t) \propto k^4$
as predicted by PLT because the eigenvalues, for $|\bk|\lesssim k_N$,
goes as $\varepsilon(\bk)\simeq 1-\al(\bhk)k^2/k_N^2$, where
$\al(\bhk)$ is a function which depends on the particular lattice (for
more details see \cite{marcos_06}).

\begin{figure}
\includegraphics[width=0.48\textwidth]{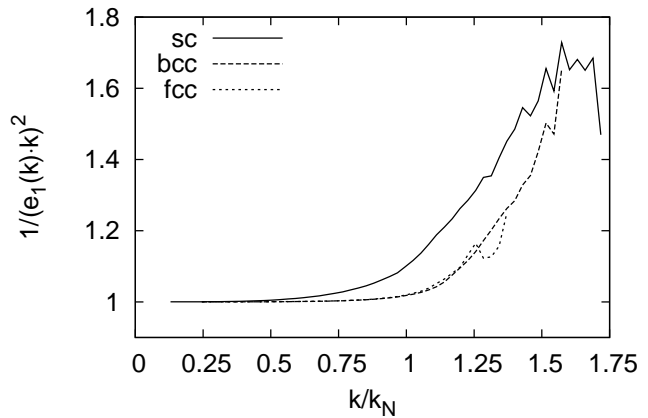}
\caption{Deviation of the polarization of the eigenvectors from the
fluid limit $\langle \tilde D_{\rm aniso}(\bk,t\gg t_0)\rangle$.
 \label{sym}}
\end{figure}

Another method to quantify explicitly the breaking of isotropy
consists in measuring the deviation of the eigenvalues corresponding
to the {\em optical} branch with the polarization of them in the fluid limit
in the direction $\hat\bk$ (see \cite{marcos_06,joyce_07}). We can
quantify it using the expression:
\be
\tilde D_{{\rm aniso}}(\bk,t)=\frac{|\tbu(\bk,t)-\hat\bk\cdot\tbu(\bk,t)\hat\bk|^2}{|\hat\bk\cdot\tbu(\bk,t)|^2}.
\ee
In the infinite realizations limit, assuming that the IC have been
set-up using the ZA [i.e. the expression \eqref{g-ZA} holds], using
Eq.~\eqref{amunu} we have:
\be
\label{aniso-average}
\langle \tilde D_{\rm aniso}(\bk,t)\rangle_{ZA}=\left|\frac{{\cal A}_{\mu\nu}{\cal
A}_{\mu\si}\hat\bk_\nu\hat\bk_\si}{({\cal A}_{\mu\nu}\hat
\bk_\mu\hat\bk_\nu)^2}\right|.  \ee
For sufficiently long times (i.e. some dynamical times $\tau_{{\rm dyn}}=1/\sqrt{4\pi G \rho_0}$) the expression
\eqref{aniso-average} is independent on how the IC have been set-up and
on the cosmological model. It depends only on the eigenvectors, i.e.,
the particular lattice:
\be
\langle \tilde D_{\rm aniso}(\bk,t\gg t_0)\rangle\approx\frac{1}{(\hat {\mathbf  e}_1(\bk)\cdot \hat\bk)^2},
\ee
where $\hat {\mathbf e}_1(\bk)$ is the eigenvector corresponding to
the optical branch, i.e., the one with maximal associated
eigenvalue. We plot this quantity in Fig.~\ref{sym}. Once again, we
see that the bcc and fcc lattices are very similar, while the breaking
of isotropy of the sc lattice is much larger.

\section{Conclusions and perspectives}

In this technical paper we have explained step-by-step how to apply
the Particle Linear Theory to a cubic Bravais lattice (sc, bcc and fcc
lattices) in a {\em cubic} simulation box. We use FFT techniques to
speed-up the numerical computations, which permits to compute the
evolution of the position of a large number of particles in a small
computation time even with modest computer resources. We have
illustrated the method computing the discreteness effects --- in the
linear regime --- resulting from the evolution of continuous density
field discretized using a perturbed bcc, fcc and sc lattice. Attending
to the tests we have performed, the bcc and fcc discretizations
present less discreteness effects --- in this regime --- than the sc
one, presenting small differences between them. They might be
therefore better choices to set-up the IC in cosmological N-body
simulations.

As pointed-out in the introduction, an important motivation of this work ---
and the reason for which we have actually developed it --- is the
study of the discreteness effects in the highly-non linear
(non-perturbative) regime. A way to estimate the discreteness effects
in this regime consists in running a set of simulations set-up with
different Bravais cubic lattices \cite{marcos_07}. They lead to
results which differ between them in a few per cent in the PS. From
the IC and the final measured PS, a lot ingredients enter in the game:
the parameters of numerical integration --- strategy and accuracy in
the computation of the force, smoothing, time-step --- , finite-size
effects, noise of the estimator, statistical fluctuations\dots It is
important to have an analytic tool to check that the differences
observed in the simulations corresponds actually to discreteness
effects. The PLT plays this role in the linear regime of gravitational
clustering (i.e. ``small'' $k$ in the PS), which strongly suggests that
the effects observed are actually discreteness ones in the {\em whole}
range of $k$.

We have described the method for an EdS universe. It is possible to
genereralize the treatment for a flat background model with
cosmological constant --- which is the currently most favored one
---, without much extra-numerical cost. An implementation of the PLT
for this kind of background model will be presented in a forthcoming
paper.

 We have considered only {\em cubic} Bravais lattices. The method can
be also applied to {\em any} Bravais lattice with the caveat that, in
some cases, the simulations box could not always be a cube, i.e., the
vectors $\bR$ might not be translated into a cube using the
transformation \eqref{translate-cubic}.

\acknowledgments{ I thank A.~Gabrielli, S.~Gaudio, J.~Lorenzana and
M.~Joyce for many helpful discussions and comments.

\appendix

\section{Ewald sum of the dynamical matrix}
\label{ewald-app}

The Ewald sum for the dynamical matrix is:
\begin{equation}
\label{dynamical_ewald2}
\mathcal{D}(\bR)=\mathcal{D}^{(r)}(\bR)+\mathcal{D}^{(k)}(\bR)
\end{equation}
with
\begin{widetext}
\begin{eqnarray}
\label{dynamical_ewald_linear_r}
&&\mD_{\mu\nu}^{(r)}(\bR\ne\mathbf0)=-Gm\sum_{{\bn}}\left[\frac{(\bR-{\bn\cdot}\bA)_\mu (\bR-{\bn\cdot}\bA)_\nu}{|\bR-{\bn\cdot}\bA|^2}\right]\frac{4\alpha^3}{\sqrt{\pi}}\exp(-\alpha^2|\bR-{\bn\cdot}\bA|^2)\\
&+&Gm\sum_{\bn}\left[\frac{\delta_{\mu\nu} }{|\bR-{\bn\cdot}\bA|^3}-3\frac{(\bR-{\bn\cdot}\bA)_\mu (\bR-{\bn\cdot}\bA)_\nu}{|\bR-{\bn\cdot}\bA|^5}\right]\left[\rm{erfc}(\alpha|\bR-{\bn\cdot}\bA|)+\frac{2\alpha}{\sqrt{\pi}}\exp(-\alpha^2|\bR-{\bn\cdot}\bA|^2)|\bR-{\bn\cdot}\bA|\right]\nonumber
\end{eqnarray}
\end{widetext}
and 
\begin{equation}
\label{dynamical_ewald_linear_k}
\mD_{\mu\nu}^{(k)}(\bR)=\frac{4\pi G m}{V_B}\sum_{\bk\ne0}\frac{1}{|\bk|^2}\exp\left(-\frac{|\bk|^2}{4\alpha^2}\right)\cos\left(\bk\cdot\bR\right)k_\mu k_\nu.
\end{equation}
Note that the sum in Eq.~\eqref{dynamical_ewald_linear_k} is over {\em
all} Fourier space and not only in the FBZ. The $\bR=0$ term is 
\be
\label{Ddiag}
\mD(\bR=\mathbf 0)=-\sum_{\bR\ne\mathbf 0,{\bn}} \mD(\bR+{\bn\cdot}\bA).
\ee
In order to sum over a minimal number of vectors in real and Fourier space we take $\alpha\approx 2.067$.

\section{Solution of the mode equations}
\label{mode-sol}

 We choose the  solutions $U_n(\bk,t)$ and $V_n(\bk,t)$ of the mode equation \eqref{modes-exp}, without any
loss of generality, satisfying 
\bse
\label{mode-boundary}
\begin{align}
&U_n(\bk,t_0)=1,\qquad \dot U_n(\bk,t_0)=0,\\
&V_n(\bk,t_0)=0,\qquad \dot V_n(\bk,t_0)=1.
\end{align}
\ese 

For an EdS universe the scale factor is \cite{peebles_80}:
\be
\label{a-EdS}
a(t)=\left(\frac{t}{t_0}\right)^{2/3}.
\ee
In this particular case the functions $U_n(\bk,t)$ and $V_n(\bk,t)$
can be calculated analytically and are:
\bse
\label{uv-exp}
\begin{align}
U_n(\bk,t)=&\tilde\al(\bk)
\left[\al_n^{+}(\bk)\left(\frac{t}{t_0}\right)^{\al_n^{-}(\bk)}+
\al_n^{-}(\bk)\left(\frac{t}{t_0}\right)^{-\al_n^{+}(\bk)}\right]\\
V_n(\bk,t)=&\tilde\al(\bk)t_0
\left[\left(\frac{t}{t_0}\right)^{\al_n^{-}(\bk)}-\left(\frac{t}{t_0}\right)^{-\al_n^{+}(\bk)}\right]
\end{align}
\ese
where
\be
\tilde\al(\bk)=\frac{1}{\al_n^{-}(\bk)+\al_n^{+}(\bk)}
\ee
and
\bse
\begin{align}
&\al_n^{-}(\bk)=\frac{1}{6}\left[\sqrt{1+24\varepsilon_n(\bk)}-1\right],\\
&\al_n^{+}(\bk)=\frac{1}{6}\left[\sqrt{1+24\varepsilon_n(\bk)}+1\right].
\end{align}
\ese
If $\varepsilon_n(\bk)>0$ the solution presents a power-law amplification mode
and a power-law decaying mode. If $-1/24<\varepsilon_n(\bk)<0$, there are two
decaying modes. Finally, if $\varepsilon_n(\bk)\leq-1/24$, the solution is
oscillatory and can be written as
\bse
\label{uv-exp-neg}
\begin{align}
U_n(\bk,t)=&\left(\frac{t}{t_0}\right)^{-\frac{1}{6}}\cos\left[\ga_n(\bk)\ln\left(\frac{t}{t_0}\right)\right]\\\nonumber
&+\frac{1}{6\ga_n(\bk)}\left(\frac{t}{t_0}\right)^{-\frac{1}{6}}\sin\left[\ga_n(\bk)\ln\left(\frac{t}{t_0}\right)\right],\\
V_n(\bk,t)=&\frac{t_0}{\ga_n(\bk)}\left(\frac{t}{t_0}\right)^{-\frac{1}{6}}\sin\left[\ga_n(\bk)\ln\left(\frac{t}{t_0}\right)\right]
\end{align}
\ese
where
\be
\ga_n(\bk)=\frac{1}{6}\sqrt{|24\varepsilon_n(\bk)+1|}.
\ee
The evolution of the displacement field from any initial
state $\bu(\bR,t_0)$ is then given by the transformation 
\be
\label{eigen_evol}
\bu(\bR,t)=\frac{1}{N}\sum_{\bk}\left[\mathcal{P}(\bk,t)\tbu(\bk,t_0)+\mathcal{Q}(\bk,t)\dot\tbu(\bk,t_0)\right]e^{i\bk\cdot\bR}
\ee
where the matrix elements of the ``evolution operators'' $\mathcal{P}$ and $\mathcal{Q}$ are
\bse
\label{evol_operators}
\begin{align}
\mP_{\mu\nu}(\bk,t)=&\sum_{n=1}^3 U_n(\bk,t)(\mathbf{\hat e}_n(\bk))_\mu(\mathbf{\hat e}_n(\bk))_\nu,\\
\mQ_{\mu\nu}(\bk,t)=&\sum_{n=1}^3 V_n(\bk,t)(\mathbf{\hat e}_n(\bk))_\mu(\mathbf{\hat e}_n(\bk))_\nu.
\end{align}
\ese

\section{A brief summary of the FFT technique}
\label{app-FFT}

Let us consider for sake of simplicity the one-dimensional
FT $\tilde f$ of the function $f$:
\be
\label{FT-1d}
\tilde f_k=\sum_{j=0}^{N-1} e^{i2\pi j k/N} f_i.
\ee
Because of the symmetries of the FT it is possible to divide the sum
\eqref{FT-1d} (with $N$ terms) into two sums with $N/2$ terms (this is
called the {\em Danielson-Lanczos} lemma):
\bea
\label{dac}
\nonumber
\tilde f_k&=&\sum_{j=0}^{N/2-1} e^{i 2\pi j k/(N/2)} f_{2j} \\
&& + e^{i 2\pi k/N}\sum_{j=0}^{N/2-1} e^{i2\pi j k/(N/2)} f_{2j+1},
\eea
i.e., an ``even'' and ``odd'' term. Therefore, at this stage, it is
possible to compute the even and odd sums at the same time, and then
sum the result to obtain the desired FT. It involves a total of
$N\times(N/2)+1$ operations, instead of $N^2$. For a number of
particles which is a power of two, we can perform recursively the
division \eqref{dac} $\ln_2 N$ times. Therefore the computation of the
$N$ terms $\tilde f_k$ involves only $N\ln_2 N$ operations. This is
called the {\em Cooley-Tukey FFT algorithm}. It exist other algorithms
(which we will not describe here), which can use {\em any} number $N$
of particles (and not only a power of two).


\end{document}